\documentclass[epj]{svjour}
\usepackage{graphicx}
\usepackage[german,USenglish]{babel}
\usepackage{epsf,epsfig,amsmath}

\newcommand{\be}{\begin{equation}}
\newcommand{\ee}{\end{equation}}
\newcommand{\ber}{\begin{eqnarray}}
\newcommand{\eer}{\end{eqnarray}}

\def\fm3{fm$^{-3}$}

\begin{document}
\begin{onecolumn}
\title{Deuteron form factors in chiral effective theory:
  regulator-independent results and the role of two-pion exchange}
\author{Manuel Pav\'on Valderrama \inst{1} 
            \thanks{\emph{Email:} m.pavon.valderrama@fz-juelich.de}
            \and Andreas Nogga \inst{1}
            \thanks{\emph{Email:} a.nogga@fz-juelich.de}
            \and Enrique Ruiz Arriola \inst{2}
            \thanks{\emph{Email:} earriola@ugr.es}
             \and Daniel~R.~Phillips \inst{3}
             \thanks{\emph{Email:} phillips@phy.ohiou.edu}
}                     
%
%
\institute{Institut f\"ur Kernphysik, Forschungszentrum J\"ulich, 52425 J\"ulich, Germany
        \and Departmento de F\'{\i}sica At\'omica, Molecular y
                Nuclear, Universidad de Granada, E-18071 Granada, Spain
      \and   Department of Physics and Astronomy, Ohio
                University, Athens, OH 45701}
\authorrunning{Manuel Pav\'on Valderrama {\it et al}}
\titlerunning{Deuteron form factors in chiral effective theory}
\date{Received: date / Revised version: date}
%
\abstract{
We evaluate the deuteron charge, quadrupole, and magnetic form factors
using wave functions obtained from chiral effective theory ($\chi$ET)
when the potential includes one-pion exchange, chiral two-pion
exchange, and genuine contact interactions. We study the manner in
which the results for form factors behave as the regulator is removed
from the $\chi$ET calculation, and compare co-ordinate- and
momentum-space approaches. We show that, for both the LO and NNLO
chiral potential, results obtained by imposing boundary
conditions in co-ordinate space at $r=0$ are equivalent to the
$\Lambda \rightarrow \infty$ limit of momentum-space calculations. The
regulator-independent predictions for deuteron form factors that
result from taking the $\Lambda \rightarrow \infty$ limit using the LO
$\chi$ET potential are in reasonable agreement with data up to
momentum transfers of order 600 MeV, provided that phenomenological
information for nucleon structure is employed. In this range the use
of the NNLO $\chi$ET potential results in only small changes to the LO
predictions, and it improves the description of the zero of the charge
form factor.
} 
\maketitle

\section{Introduction}
\label{sec-intro}

Elastic electron scattering from the deuterium nucleus, the deuteron,
has long been used as a means to assess the reliability of different
nuclear forces. For recent reviews of experimental and theoretical
work on the elastic electron-deuteron reaction see
Refs.~\cite{GG01,vOG01,Si01}.  In this reaction the ability to vary
the momentum transfer, $q$, to the nucleus means the electromagnetic
structure of the deuteron can be probed on a variety of different
scales. Indeed, in the ``impulse approximation''---where two-nucleon
currents are neglected---the matter distribution inside the nucleus is
imaged, thereby providing a direct test of the $NN$
interaction used to predict it.  In this paper we examine elastic
electron-deuteron scattering for $q < 1$ GeV, where we anticipate that
the deuteron's electromagnetic structure will be governed by the
spontaneously and explicitly broken chiral symmetry of QCD.

This feature of QCD plays a role even though quark degrees of freedom
are not resolved at this momentum scale. Indeed, confinement
guarantees that for $q < 1$ GeV it is useful to employ an effective
field theory (EFT) in which nucleons and pions (and possibly 
baryon resonances) are the explicit degrees of freedom. Here we will
use chiral perturbation theory ($\chi$PT) which is an EFT in which the
nucleons and pions interact with each other in a manner consistent
with the chiral symmetry of QCD and the pattern of its
breaking. $\chi$PT does not aim to describe hadronic and nuclear
structure down to arbitrarily short distances, but is based on the
inability of low-momentum-transfer probes to resolve the details
of short-range interactions.  It is thus ideally suited to the
treatment of deuteron electromagnetic structure in the regime $q <
1$ GeV.

Chiral perturbation theory organizes the Lagrangian that describes
nucleon and pion (and photon) interactions in powers of the small
expansion parameter $P \equiv \frac{m_\pi,p}{\Lambda_{\chi \rm SB}}$,
where $\Lambda_{\chi \rm SB} \sim m_\rho, 4 \pi f_\pi$ represents the
scale of chiral-symmetry breaking, which is indicated by the $\rho$ mass
and the pion decay constant, while $p$ is the momentum of the particles
  involved, and $m_\pi$ the pion mass. This theory has had
  considerable success in describing pion-nucleon and meson-meson
  interactions~\cite{Bernard:2006gx,Scherer:2005ri}. Weinberg proposed
  to extend $\chi$PT to few-nucleon systems by making a $\chi$PT
  expansion for the nucleon-nucleon interaction, $V$, and then taking
  the non-perturbative character of nuclear systems into account by
  solving the Schr\"odinger equation based on this
  potential~\cite{We90,We91,We92}. The result is quantum mechanics
  with fixed particle number for few-nucleon systems in which the
  potential (and other operators) has a $\chi$PT expansion. This has
  been dubbed chiral effective theory ($\chi$ET). This approach is
  reviewed in
  Refs.~\cite{Beane:2000fx,Bedaque:2002mn,Epelbaum:2005pn,Phillips:2007bw}.

Recently there has been considerable controversy about the correct way
to implement this
approach~\cite{PavonValderrama:2005uj,Nogga:2005hy,Epelbaum:2006pt,Birse:2007sx}. It
has been argued that Weinberg's original power counting needs to be
modified for the short-range or contact interaction piece of $V$ if
one wishes to obtain regulator-independent predictions over a wide
range of cutoffs, $\Lambda$.  However, as shown already in
Ref.~\cite{Beane:2001bc}, in the ${}^3$S$_1$--${}^3$D$_1$ channel the
leading-order chiral potential can be renormalized using the single
contact interaction it contains. The limit $\Lambda \rightarrow
\infty$ can be taken, as has since been confirmed by several
authors~\cite{Nogga:2005hy,PavonValderrama:2005gu,Birse:2007sx,Yang:2007hb}. Thus
the discussions of
Ref.~\cite{PavonValderrama:2005uj,Nogga:2005hy,Epelbaum:2006pt,Birse:2007sx}
are not relevant to our results for leading-order (LO) $\chi$ET, and
we will not contribute to that discussion here.

However, the situation is not as clear once sub-leading corrections
are included in the $NN$ interaction $V$. In particular, while one
naively expects that these corrections to $V$ can be treated in
perturbation theory, previous works have included them
non-perturbatively, by solving the Schr\"odinger equation with the
full $V$ at a fixed order in $\chi$PT. In such a calculation the sub-leading
contributions to $V$ dominate over the leading-order part of the
$\chi$PT potential for $r \ll 1/m_\pi$, and the singularity of $V$ at short
distances therefore becomes more severe as the chiral order to which $V$ is
computed is increased. In what follows we adopt such a
non-perturbative treatment of
$V$, even though it leads to a paradox when the next-to-leading-order
[$O(P^2)$] $V$ is employed.  In standard $\chi$PT, where only nucleons
and pions are taken as the relevant degrees of freedom, the
next-to-leading order (NLO) $V$ calculated for $r \sim 1/m_\pi$ is
repulsive for $r \ll 1/m_\pi$. As is well known from the theory of
singular potentials~\cite{Case:1950an}, this means that the spectrum
of the potential is predicted in the limit that the cutoff on the
potential is removed, i.e. that the $\chi$PT potential is taken to be
valid for all $r$. In consequence in the limit $\Lambda
\rightarrow \infty$ there is no opportunity to tune the contact
interaction and thereby reproduce the binding energy of the shallow $NN$ bound
state in the ${}^3$S$_1$--${}^3$D$_1$
channel~\cite{PavonValderrama:2005wv}.

Therefore, we can only present results for the LO [$O(P^0)$] and
next-to-next-to-leading order [NNLO=$O(P^3)$] chiral potentials
$V$. Both of those potentials are attractive for $r \ll 1 /
m_\pi$, which means that the inclusion of contact interaction(s) is
mandatory if cutoff-independent predictions are to be
obtained. Moreover, at these orders, and in the
${}^3$S$_1$--${}^3$D$_1$ channel of $NN$ scattering, there is no
problem with Weinberg's original proposal: the three short-distance
$NN$ operators his $\chi$PT expansion of $V$ predicts should be
present at NNLO are sufficient to renormalize the attractive
long-range potential generated by pion-nucleon dynamics at $r \sim
1/m_\pi$.

In Section~\ref{sec-randpspace} we compute the deuteron wave functions
obtained from the LO and NNLO chiral potentials $V$. We show that when
we compute using a momentum-space cutoff on $V$ but demand that the
deuteron binding energy be reproduced we obtain cutoff-independent
predictions for the radial wave functions $u(r)$ and $w(r)$ for $r \gg
1/\Lambda$. In the limit $\Lambda \rightarrow \infty$ the result of
such a momentum-space calculation then agrees with that obtained via
the co-ordinate-space approach advocated in
Refs.~\cite{PavonValderrama:2005gu,PavonValderrama:2005wv}. We use existing wave functions to
make this demonstration for the LO $\chi$PT potential $V$, and
we also obtain, for the first time, 
NNLO momentum-space wave functions in $\chi$ET for
$\Lambda > 1$ GeV.  This allows us to make the connection between
momentum- and co-ordinate-space  not just at LO, but also at NNLO,
and, by extension, for any order at which $V$ is attractive at short
distances. 

These wave functions, while not directly observable, do enter in the
matrix elements of nuclear current operators that are probed in
electron scattering from the deuterium nucleus.  In particular, to the
extent that the impulse approximation is valid, the form factors that
determine all elastic electron-deuteron scattering observables in the
one-photon-exchange approximation can be expressed as Bessel
transforms of bilinears of these wave functions.  We present results
for these form factors in Sec.~\ref{sec-LOFF}.  These results are
based on a chiral expansion of the deuteron current operators
\cite{Rho:1990cf,Park:1995pn}. Up to relative order $P^3$ there are no
two-nucleon contributions in either the charge operator, $J^0$ or the
vector-part of the current, ${\bf J}$, and the impulse approximation
holds.  The deuteron form factors have already been calculated in such
an approach (including two-body currents) in
Refs.~\cite{Phillips:1999am,Walzl:2001vb,Phillips:2003jz,Phillips:2006im}.
Here, we follow \cite{Phillips:2003jz} and focus on tests of deuteron
structure by computing the ratio $G_C/G_E^{(s)}$, where the
denominator is the nucleon isoscalar form factor. We show that cutoff
artifacts disappear in this ratio as $\Lambda \rightarrow \infty$,
allowing us to obtain the first regulator-independent, LO, results for
$G_C$ in the literature. We find that if we use phenomenological input
for $G_E^{(s)}$ these results are in good agreement with data for
$G_C$.

If $\chi$ET is to be a systematic expansion for $NN$ interactions we
should be able to also obtain results at higher orders in the $P$
expansion. Therefore in Sec.~\ref{sec-higho} we compute the results
found for form factors when $G_C$, $G_Q$, and $G_M$ are computed with
deuteron wave functions found using the NNLO $\chi$PT potential.  This
potential includes various contributions from two-pion-exchange
mechanisms.  Once again we find that the cutoff artifacts in
observables vanish as $\Lambda \rightarrow \infty$.  Importantly, we
find that the shift from LO to NNLO results for the form factors is
small for $q < 600$ MeV.

This suggests that if we {\it could} do a NLO calculation of deuteron
structure it would be close to the regulator-independent LO result in
this kinematic domain. In the absence of a calculation at NLO with
$\Lambda \rightarrow \infty$ in the theory with only explicit nucleon
and pion degrees of freedom this is purely supposition. Nevertheless,
we believe that our results with the NNLO $V$ and $\Lambda \rightarrow
\infty$ give realistic estimates for two-pion-exchange contributions
to the form factors. We hold this view because the $O(P^3)$ pieces of
$V$ that make the difference between an NNLO and NLO calculation are
numerically much more important than NLO
pieces~\cite{Ordonez:1995rz,Kaiser:1997mw,Epelbaum:1999dj}. This is in
large part because our NNLO $V$ includes the most important
contributions of the $\Delta(1232)$ to the $NN$
interaction~\cite{Ordonez:1995rz,Kaiser:1998wa,Epelbaum:1999dj}. In
particular, if a consistent power counting for nuclear interactions
requires taking the $\Delta(1232)$ into account as an explicit degree
of freedom \cite{Ordonez:1995rz,Kaiser:1998wa,Krebs:2007rh} our NNLO
calculation will produce results similar to an NLO calculation in the
theory with explicit Deltas.

The calculation presented in Sec.~\ref{sec-higho} shows significant
sensitivity in the position of $G_C$'s zero to the details of the NNLO
$\chi$PT potential. However, any attempt to use data on the $q_0$ for
which $G_C(q_0)=0$ to pin down the two-pion-exchange part of the $NN$
force is clouded by the presence of two-nucleon pieces of $J_0$
at $O(P^3)$---the same order relative to leading as that at which the
two-pion-exchange pieces of $V$ first occur. Similar ambiguities
bedevil attempts to use other features of deuteron electromagnetic
form factors to identify aspects of the $NN$ force beyond one-pion
exchange.  We provide an estimate of the impact on our results of
$O\left(\frac{P^3}{\Lambda^2 M}\right)$ and $O\left(\frac{P^2}{M^2}\right)$, pieces of $J_0$ in
Sec.~\ref{sec-higho}. But a full calculation of their contribution
when $\Lambda \rightarrow \infty$ is beyond the scope of this paper.
In Sec.~\ref{sec-concl}, we summarize our findings and conclude.

\section{$\chi$ET wave functions: equivalence of
  co-ordinate- and momentum-space solutions}

\label{sec-randpspace}

$\chi$ET for few-nucleon systems results when $\chi$PT is applied to
derive potentials and current operators which can then be used in a
non-relativistic quantum-mechanical (fixed-particle-number)
framework. Corrections to the static-nucleon picture may be treated in
such a framework, e.g.  here we incorporate the $1/M$ corrections to
$V$ as given by \cite{Kaiser:1997mw}.  $1/M^2$ pieces of $V$ and the
effects of higher-Fock-space (e.g. $\pi NN$) states can also be
incorporated, but will not concern us significantly here, although the
former will be briefly discussed in Sec.~\ref{sec-higho}.

\subsection{Regulator-independent results for deuteron wave functions
  with the leading-order $\chi$PT potential}

As first discussed by
Weinberg~\cite{We90,We91} the leading-order potential in the $\chi$ET
reproduces the time-honoured one-pion exchange (OPE) potential
\begin{eqnarray}
V^{(0)}(\vec{p},\vec{p}')  &=& -\frac{g_A^2}{4 f_\pi^2} \tau_1
\cdot \tau_2 \frac{{\bf \sigma}_1 \cdot ({\bf p}' - {\bf p}) \  {\bf \sigma}_2 \cdot
({\bf p}' - {\bf p})}{({\bf p}' - {\bf p})^2 + m_\pi^2} + C_S + C_T \ 
{\bf \sigma}_1 \cdot {\bf \sigma}_2 
\label{eq:LOV}
\end{eqnarray}
We can now solve the Schr\"odinger equation in its momentum-space form,
i.e. the homogeneous Lippmann-Schwinger equation: 
\begin{equation}
\label{eq:schroedinger}
\langle {\bf p}|\psi_{\rm LO} \rangle_\Lambda=
G_0(p)\int   \frac{\hbox{d}^3 p'}{(2 \pi)^3} \,
V^{(0)}_\Lambda(\vec{p},\vec{p}') \langle {\bf p}'|\psi_{\rm LO} \rangle_\Lambda~,
\end{equation} 
where $\Lambda$ is the scale at which the potential $V$ is regulated,
and $G_0(p)=(-B_d-p^2/M)^{-1}$ is the (free, center-of-mass frame)
two-nucleon propagator, with $B_d$ and $M$ denoting the deuteron
binding energy and nucleon mass, respectively.  The
OPE contribution is determined
through the pion mass $m_\pi = 138$~MeV, the axial coupling constant (for which
we adopt $g_A=1.29$) and the pion-decay constant $f_{\pi}=92.4$~MeV. 
In the triplet channel, only one linear combination of 
the contact interaction parameters, $C_S + C_T$, is relevant and is 
adjusted for each cutoff $\Lambda$ to reproduce the deuteron binding 
energy $B=2.225$ MeV as previously done, e.g., in \cite{Nogga:2005fv}.
We perform the regularization using exponential cutoff functions  
\begin{equation}
V^{(0)}_\Lambda(\vec{p},\vec{p}') = \exp \left( - \, { p^8 \over \Lambda^8}\right) \ V^{(0)} (\vec{p},\vec{p}') \exp  \left( - \, { {p'}^8 \over \Lambda^8}\right).
\end{equation}

Alternatively, the interaction can be Fourier-transformed to
co-ordinate-space resulting in
\begin{eqnarray}
V ( {\bf r} ) &=& \left[ \tilde C_S  + \tilde C_T  \sigma_1
\cdot \sigma_2 \right] \delta^{(3)} ( {\bf r}) +  \frac{g_A^2}{2 f_\pi^2}
\, \sigma_1 \cdot {\vec \nabla} \sigma_2 \cdot {\vec \nabla} \,
\frac{e^{-m_\pi r}}{r}.
\label{eq:LOV-coor}
\end{eqnarray} 
The Fourier transformation uniquely determines the long-range part of
the interaction. (Note, however, that it involves a redefinition of
the contact interactions to absorb a piece of $V^{(0)}$ that is a
constant in momentum space.)  The deuteron wave function in the pn CM
system can be represented as
\begin{eqnarray}
\Psi (\vec r) &=& \frac1{\sqrt{4\pi}r } \Big[ u(r) \sigma_p \cdot
\sigma_n + \frac{w(r)}{\sqrt{8}} \left( 3 \sigma_p
\cdot \hat r \, \sigma_n \cdot \hat r - \sigma_p \cdot \sigma_n \right)
\Big] \chi_{pn}^{s m_s}
\label{eq:deut-coor}
\end{eqnarray} 
with the total spin $s=1$ and $m_s=0,\pm 1$ and $\sigma_p$ and
$\sigma_n$ the Pauli matrices for the proton and the neutron,
respectively. The functions $u(r)$ and $w(r)$ are the radial
$^3$S$_1$ and $^3$D$_1$ components of the relative wave function,
respectively.  

Even if we consider only $r > 0$ and ignore the presence
of the three-dimensional delta functions, this is a singular
quantum-mechanical potential. In the absence of any short-distance
regulator, the resulting Hamiltonian is unbounded from below  
and the spectrum contains an infinite number of bound states. This happens for
any value of the parameters. On the other hand, it is known that 
physically there is only one bound $NN$ state: the deuteron. 
The presence of these unphysical, spurious bound states should 
not have any impact on low-energy observables, including the form factors 
at low momentum transfers. We will come back to this issue in Section~\ref{sec-LOFF}. 

To derive the wave function from Eq.~(\ref{eq:LOV-coor}), we use
the representation of Eq.~(\ref{eq:deut-coor}) and find the usual coupled 
one-dimensional differential equations for the radial
wave functions $u$ and $w$ in the presence of a tensor potential:
\begin{eqnarray}
\label{eq:deuterondiffeq}
-u''(r)+U_s(r)u(r)+U_{sd}(r) w(r)&=&-\gamma^2 u(r),
\nonumber\\
-w''(r)+U_{sd}(r)u(r)+\biggl[U_d(r)+\frac{6}{r^2}\biggr]w(r)&=&
-\gamma^2 w(r),
\label{eq:SErspace}
\end{eqnarray}
The coupled-channel reduced potential ($ U = 2 \mu_{pn} V$
with $2 \mu_{pn} = 2 M_p M_n /(M_p+M_n)$)  is given by
\begin{equation}
U_s=U_C,\qquad U_{sd}=2\sqrt{2}U_T,\qquad U_d=U_C-2U_T,
\end{equation}
with 
\begin{eqnarray}
U_C&=&-\frac{m_\pi^2 M g_A^2}{16\pi f_\pi^2}\frac{e^{-m_\pi r}}{r}~,
\nonumber\\
U_T&=&-\frac{m_\pi^2 M g_A^2}{16\pi f_\pi^2}\frac{e^{-m_\pi
    r}}{r}\biggl(1+\frac{3}{m_\pi r}
+\frac{3}{(m_\pi r)^2}\biggr)~,
\label{eq:LOpotrspace}
\end{eqnarray}
provided that $r > 0$.

For our co-ordinate-space calculations, the equations (\ref{eq:SErspace}) 
are solved subject to the boundary conditions 
\begin{eqnarray}
\label{eq:longdist}
u(r)&\rightarrow& A_S \,e^{-\gamma r}~,
\nonumber\\
w(r)&\rightarrow& \eta \,A_S \,e^{-\gamma r}\biggl(1+\frac{3}{\gamma
  r}+\frac{3}{(\gamma r)^2}\biggr)~,
\end{eqnarray}
for  $r \rightarrow \infty$. Here, $\gamma=\sqrt{M B_d}$ is 
the deuteron wave number, $A_S$ is a 
normalization constant and can be chosen to  guarantee that
\begin{equation}
\int_0^\infty \hbox{d}r\bigl( u^2(r) +w^2(r)\bigr)=1~,
\end{equation}
and $\eta$ is the asymptotic D/S ratio.

The form of the singularity of the OPE potential at short-distances 
implies that, for sufficiently small $r$, the components $u$ and $w$ 
are given by ~\cite{Be01,PavonValderrama:2005gu,Sp00}: \begin{eqnarray}
\label{eq:asymptoticdeuteron}
u_{short}(r)&=&A_S \frac{1}{\sqrt{3}}
\biggl(\frac{r}{R}\biggr)^{3/4}\biggl[-C_{2R}e^{-4\sqrt{2}\sqrt{R/r}}
+2^{3/2}|C_{2A}|\cos\left(4\sqrt{\frac{R}{r}}+\phi\right)\biggr],
\nonumber\\ w_{short}(r)&=&A_S\frac{1}{\sqrt{3}}
\biggl(\frac{r}{R}\biggr)^{3/4}\biggl[\sqrt{2}C_{2R}e^{-4\sqrt{2}\sqrt{R/r}}
+2|C_{2A}|\cos\left(4\sqrt{\frac{R}{r}}+\phi\right)\biggr].
\end{eqnarray} 
$C_{2A}$ and $C_{2R}$ are normalization constants which have been
determined in \cite{PavonValderrama:2005gu}, $R$ is a new length scale that enters the
non-perturbative problem, given by $R={\textstyle \frac{3
    g_A^2 M}{32 \pi f_\pi^2}}$. When
Eqs.~(\ref{eq:SErspace})--(\ref{eq:LOpotrspace}) are solved in
Ref.~\cite{PavonValderrama:2005gu} the phase $\phi$ is determined by the boundary
condition at $r=0$, and so $\phi$ is regulator independent, and is a
function only of the scales $m_\pi$, $\gamma$, and $R$.  

The wave functions $u$ and $w$ can now be calculated by applying
standard techniques to 
Eqs.~(\ref{eq:SErspace})--(\ref{eq:LOpotrspace}) and employing the
long-distance boundary conditions
(\ref{eq:longdist})~\cite{PavonValderrama:2005gu}.  The asymptotic
D-to-S ratio $\eta$ is then determined by the requirement that the
numerical solution matches the short-distance behavior
Eq.~(\ref{eq:asymptoticdeuteron}). In practice, this is done by
imposing the additional boundary condition 
$u(r_c)-\sqrt{2}w(r_c)=0$,
which is a direct
consequence of Eq.~(\ref{eq:asymptoticdeuteron}), at a value of $r_c
\ll R$. For this purpose we choose
$r_c=0.1$--0.2 fm~\cite{PavonValderrama:2005gu}, and find 
$\eta_{\rm OPE} (\gamma ) \approx 0.026333$.

The necessity to choose a finite $r_c$ leads to a small numerical
uncertainty of the constants $C_{2A}$, $C_{2R}$ quoted in
Ref.~\cite{PavonValderrama:2005gu}.  These uncertainties can, however,
be made arbitrarily small. Furthermore, any deuteron matrix element
that is finite when computed with the wave functions $u$ and $w$
obtained in this way is insensitive to $r_c$ as long as $r_c$ is taken
small enough~\cite{PlatterPhillips}.  Therefore, for all practical
purposes, the wave functions found by the technique of
Ref.~\cite{PavonValderrama:2005gu} are the deuteron wave functions
found with a strict contact interaction. 
The contact parameter $C_S + C_T$  
has been converted to a
boundary condition that is imposed at $r=0$~\cite{vanKolck:1999}.

For non-singular interactions, it is clear that any solution obtained
in momentum space will be the Fourier-transform of the corresponding
co-ordinate-space solution. The question arises whether this holds
for the solution of the singular OPE interaction based on
Eq.~(\ref{eq:schroedinger}).  In momentum space the solution is only
ever determined for a finite cutoff $\Lambda$, and it is not {\it a
  priori} clear that the limit $\Lambda \rightarrow \infty$ of the
momentum-space wave function will, upon taking the Fourier transform,
lead to the co-ordinate-space wave function found using
Eqs.~(\ref{eq:SErspace})--(\ref{eq:LOpotrspace}).

\begin{figure}[tb]
\centerline{\includegraphics*[width=80mm,angle=0]{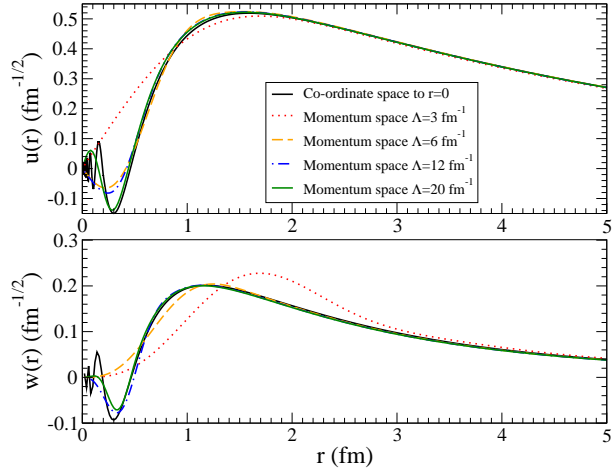}}
\caption{\label{fig-wfNoggaPVRA} Comparison of radial deuteron wave
functions $u(r)$ and $w(r)$ computed with the LO $\chi$ET potential
using a momentum-space
regularization~\cite{Nogga:2005hy,Nogga:2005fv} 
and the co-ordinate-space regularization of
Ref.~\cite{PavonValderrama:2005gu}.}
\end{figure}

In Fig.~\ref{fig-wfNoggaPVRA}, we show that, for any $r > 0$, the
momentum-space solution does indeed approach this co-ordinate-space
solution for $\Lambda \rightarrow \infty$.  It is reassuring that one
obtains equivalent results in both schemes. This confirms that fitting
the contact interactions in a momentum-space calculation corresponds
to imposing proper co-ordinate-space boundary conditions and matching
to the short-distance form of the wave functions (\ref{eq:asymptoticdeuteron}).

However, the relationship between $\Lambda$ and the numerically necessary
co-ordinate-space cutoff $r_c$ remains somewhat opaque. For theories in
which the long-range-part of the potential is absent
one can establish that a sharp cut-off
$\Lambda$ and the matching radius $r_c$ for the Schr\"odinger equation
with a boundary condition are related by $r_c = { \pi \over 2
  \Lambda}$ \cite{ERAPVM07}.  But here, given our choice of regulator
and the presence of a long-range potential (one-pion exchange) this
relation is no longer valid. Indeed, especially for $w(r)$, the wave
function is only converged for distances
\begin{equation}
r \gg \frac{\pi}{2 \Lambda}.
\end{equation}
Therefore we conclude that simple arguments for ``reasonable'' choices
of the momentum-space cutoff $\Lambda$ at which results converge to
the $\Lambda \rightarrow \infty$ limit can be misleading.

\subsection{Regulator-independent results for deuteron wave functions 
with the NNLO $\chi$ET potential}

In Sec.~\ref{sec-LOFF} we use the (co-ordinate- and momentum-space)
wave functions we have obtained thus far to predict the
electromagnetic structure of the deuterium nucleus. However, this
represents only a LO calculation in $\chi$ET, so before performing
such a calculation of deuterium electromagnetic form factors we will
explain how the two-pion exchange (TPE) contribution to the $NN$
potential $V$ is incorporated in $\chi$ET.

As outlined in the introduction, $\chi$PT can be used to expand the
long-range ($r \sim 1/m_\pi$) part of the NN potential. Here we do
this up to NNLO. The explicit expressions in momentum- and
co-ordinate-space were given in Ref.~\cite{Kaiser:1997mw}.  
Note,
however, that the momentum-space expressions of Ref.~\cite{Kaiser:1997mw}
are opposite in sign to ours, since different conventions are
used. Also, it should be noted that we have used
$g_A=1.26$ for the TPE part of the NNLO potential. 

\begin{table}[bth]
\caption{\label{tab:cival} The four different sets of values for the
  chiral coefficients from ${\cal L}_{\pi N}^{(2)}$ that we consider
  in this work.}
\begin{center}
\begin{tabular}{c|c|c|c|c}
  Set & Source & $c_1 ({\rm GeV}^{-1}) $ & $c_3 ({\rm GeV}^{-1}) $ & $c_4
({\rm GeV}^{-1}) $ \\ \hline
Set I & $\pi N$~\cite{Buettiker:1999ap}  & -0.81  & -4.69    & 3.40   \\ 
Set II & $NN $~\cite{Rentmeester:1999vw} 
& -0.76  & -5.08   & 4.70  \\ 
Set III &  $NN $~\cite{Epelbaum:2003xx}  
& -0.81  & -3.40    & 3.40   \\ 
Set IV &  $NN$~\cite{Entem:2003ft} & -0.81  & -3.20   & 5.40  \\ 
\end{tabular}
\end{center}
\end{table}

For the calculations in co-ordinate-space, we follow the formalism
introduced by two of us in Ref.~\cite{PavonValderrama:2005wv} where
two-pion-exchange effects were included non-perturbatively. For the
NNLO interaction the short-range potential behaves like
~\cite{Kaiser:1997mw,Friar:1999sj,Rentmeester:1999vw}
\begin{eqnarray} 
U_{s}^{\rm TPE} (r) &\to& \frac{R_{s}^4}{ r^6 }, \nonumber \\ 
U_{sd}^{\rm TPE} (r)  &\to& \frac{R_{sd}^4}{ r^6 }, \nonumber \\ 
U_{d}^{\rm TPE} (r)  &\to& \frac{R_{d}^4}{ r^6 };
\label{eq:TPE}
\end{eqnarray} 
where 
\begin{eqnarray} 
(R_{s})^4 &=& \frac{3 g_A^2}{128 f_\pi^4 \pi^2 } ( 4 - 3 g_A^2 + 24 \bar
c_3 - 8 \bar c_4 ), \nonumber \\ (R_{sd})^4 &=& - \frac{3 \sqrt{2}
g_A^2}{128 f_\pi^4 \pi^2 } (-4 + 3 g_A^2 - 16 \bar c_4 ), \nonumber \\
(R_{d})^4 &=& \frac{9 g_A^2}{32 f_\pi^4 \pi^2 } (-1+2 g_A^2 + 2 \bar c_3 -
2 \bar c_4 );
\label{eq:vdw_triplet}
\end{eqnarray} 
and $ \bar c_i = M c_i$ are the low energy chiral couplings (LECs) appearing
in $\pi N $ scattering \cite{Bernard:1996gq}. We will present results 
for the sets of $c_i$ parameters shown in Table~\ref{tab:cival}, which
are the same four sets as were used in 
Ref.~\cite{PavonValderrama:2005uj}. 
In contrast to OPE, this interaction yields two attractive eigenchannels 
upon diagonalization. Therefore, the short-distance wave functions in the
${}^3$S$_1$ and ${}^3$D$_1$ channels
take the form~\cite{PavonValderrama:2005wv}:
\begin{eqnarray}
u_{short} (r) &=& \left\{ C_+ \left(\frac{r}{R_+}\right)^{\frac32} \cos
\theta \sin\left[ \frac{1}{2} \frac{R_+^2}{r^2} + \phi_+\right] +
C_{-} \left(\frac{r}{R_-}\right)^{\frac32} \sin \theta
\sin\left[\frac{1}{2} \frac{R_-^2}{r^2} + \phi_-\right]
\right\}\nonumber\\ w_{short}(r) &=& \left\{ -C_+
\left(\frac{r}{R_+}\right)^{\frac32} \sin \theta \sin\left[
  \frac{1}{2} \frac{R_+^2}{r^2} + \phi_+\right] + C_{-}
\left(\frac{r}{R_-}\right)^{\frac32} \cos \theta \sin\left[\frac{1}{2}
  \frac{R_-^2}{r^2} + \phi_-\right] \right\}\nonumber\\
&&
\label{eq:uwsd}
\end{eqnarray}  
where the coefficients $C_{+}$ and $C_{-}$ and the phases $\phi_+$ and
$\phi_-$ are arbitrary. The scales $R_+$ and $R_-$ and the angle
$\theta$ are determined by diagonalizing the potential
Eq.~(\ref{eq:TPE}).  Because the short-distance wave functions
(\ref{eq:uwsd}) do not constrain the two components $u(r)$ and $w(r)$
as strongly as Eq.~(\ref{eq:asymptoticdeuteron}) does in the case of
OPE, any solution based on the Schr\"odinger equation
(\ref{eq:SErspace}) and the long-distance boundary conditions
Eq.~(\ref{eq:longdist}) can be matched to the $r \ll R_-, R_+$ wave
function. Apart from an overall normalization, the long-distance wave
function of the deuteron depends on two parameters, $\gamma$ and
$\eta$. Once these parameters have been fixed, e.g. by experiment, we
are able to calculate the wave function to arbitrarily small distances
$r>0$ uniquely and regulator-independently.  

The numerical solution can thus be used to determine the parameters of
the short-distance solution (\ref{eq:uwsd}). In this way $\gamma$ and
$\eta$ are linked to two of the contact interactions that are present
in the NNLO $\chi$PT potential. The third contact interaction does not
affect the bound state, but can be determined by an examination of
continuum wave functions. For example, for Set IV of the LECs we have
$R_+ = 2.11~{\rm fm}$ and $R_- =1.16~{\rm fm}$.  Choosing the
normalization of the long distance wave function such that $u (r) \to
e^{-\gamma r} $, and fixing $\gamma$ to its experimental value, the
dependence of $C_{+,-}$ and $\phi_{+,-}$ on $\eta$ can be
computed. The expressions can be found in
Ref.~\cite{PavonValderrama:2005wv}, where an explicit expression for
the dependence of $A_S$ on $\eta$ for the case of this potential is
also given.


Since the dominant singularity of the NNLO $\chi$PT potential is
$r^{-6}$ the solution of the ${}^3$S$_1$--${}^3$D$_1$ bound-state
problem when this potential is used for all $r$ leads to a deuteron
wave function that approaches zero more rapidly as $r \rightarrow 0$
than does the LO $\chi$ET wave function.  As a consequence, deuteron
matrix elements computed with wave functions where the NNLO potential
is treated non-perturbatively have improved ultraviolet convergence
properties~\cite{PVRA06b}. In particular, matrix elements such as
$\langle 1/r \rangle$ and $\langle 1/r^2 \rangle$, which are somewhat
sensitive to short-distance pieces of the wave function, take values
that agree quite well those found when sophisticated $NN$ potential
models are employed for their evaluation.

Now we turn our attention to momentum-space solutions of
Eq.~(\ref{eq:schroedinger}) with the NNLO TPE potential derived in
Refs.~\cite{Ordonez:1995rz,Kaiser:1997mw}. We adjust the 
momentum-independent contact interaction $C_S + C_T$ so that the experimental
deuteron binding energy is reproduced. A numerical
Fourier transform allows us to determine $\eta$ as a function of
the cutoff $\Lambda$. 
We find that for cutoffs $\Lambda \approx 3.9$~fm$^{-1}$,
8.45~fm$^{-1}$, 14.1~fm$^{-1}$ and 20.9~fm$^{-1}$, the central value
of the experimental
D/S ratio, $\eta=0.0256$~\cite{Rodning:1990zz}, 
is reproduced. Therefore at these values of the cutoff the coefficient of the
second contact interaction needed to solve the
${}^3$S$_1$--${}^3$D$_1$ bound-state problem can be chosen to be zero.

When the momentum-space problem is solved in this way and the
momentum-space wave functions are (numerically) transformed to
co-ordinate-space the long-distance behavior matches that obtained in
Ref.~\cite{PavonValderrama:2005wv}. Therefore, for these cutoffs, we
should find equivalent solutions in momentum and co-ordinate-space
in the region where $r \gg \frac{\pi}{2 \Lambda}$.

\begin{figure}[tb]
\centerline{\includegraphics*[width=80mm,angle=0]{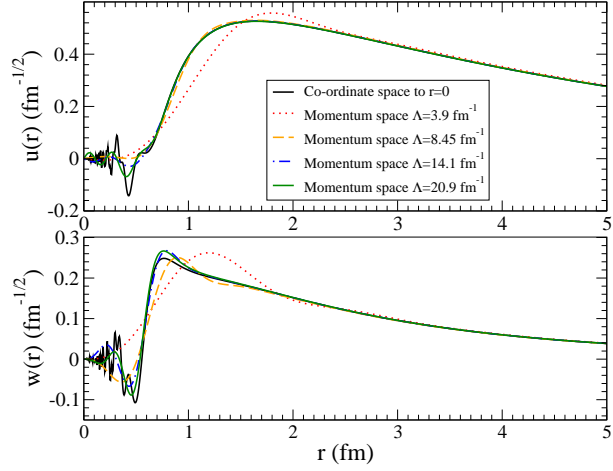}}
\caption{\label{fig-wfNoggaPVRA-TPE} Comparison of radial deuteron
  wave functions $u(r)$ and $w(r)$ computed with the NNLO $\chi$ET
  potential using a momentum-space regularization and the
  co-ordinate-space regularization of
  Ref.~\cite{PavonValderrama:2005wv}.}
\end{figure}

In Fig.~\ref{fig-wfNoggaPVRA-TPE}, we compare the Fourier-transformed
wave functions of the momentum-space calculations and the
regulator-independent result of the co-ordinate-space calculation.
For the cutoff range that we consider here, we find agreement down to
radii $r \approx 1$~fm.  We hypothesize that if $\Lambda \rightarrow
\infty$ the wave functions agree for all $r > 0$.  But as before, the
momentum-space cutoffs need to be surprisingly large to find good
agreement. E.g., for cutoff $\Lambda \approx 3.9$~fm$^{-1}$,
there are visible deviations to the r-space wave function at $r
\approx 3$~fm, and even for the largest cutoffs we observe deviations in the
peak structure of $w(r)$ around $r \approx 1$~fm. In the next section
we will test whether this deviation has any noticeable effect on
deuteron electromagnetic form factors.

\section{Deuteron form factors at leading order in $\chi$ET}

\label{sec-LOFF}

Elastic electron-deuteron scattering can be parameterized by three 
independent form factors \cite{GG01,vOG01,Si01}, which we will here write 
as Breit-frame matrix elements of
the two-nucleon four-current $J_\mu$:
\begin{eqnarray}
G_C&=&\frac{1}{3 |e|} \left(\left \langle 1\left|J^0\right|1 \right \rangle + 
\left \langle 0\left|J^0\right|0 \right \rangle + \left \langle -1\left|J^0\right|-1 \right \rangle
\right),\label{eq:GC}\\ 
G_M&=&-\frac{1}{\sqrt{2 \eta} |e|} 
\left \langle1\left|J^+\right|0\right \rangle,
\label{eq:GM}\\
G_Q&=&\frac{1}{2 |e|\eta M_d^2} 
\left(\left \langle 0\left|J^0\right|0 \right \rangle
- \left \langle 1\left|J^0\right|1 \right \rangle\right).
\label{eq:GQ}
\end{eqnarray}
The form factors defined in Eqs.~(\ref{eq:GC})--(\ref{eq:GQ}) are
related to the static moments of the nucleus by:
\begin{eqnarray}
G_C(0)&=&1,\\
G_Q(0)&=&Q_d,\\
G_M(0)&=&\mu_d \frac{M_d}{M},
\end{eqnarray}
with $M_d$ the deuteron mass, $Q_d$ the deuteron quadrupole moment,
and $\mu_d$ the deuteron magnetic moment in units of nuclear
magnetons.  In Eqs.~(\ref{eq:GC})--(\ref{eq:GQ}), we have labeled the
deuteron states by the projection of the deuteron spin along the
direction of the three-vector ${\bf p}_e' -{\bf p}_e \equiv {\bf q}$,
and $\eta \equiv Q^2/(4 M_d^2)$, with $Q^2=|{\bf q}|^2$ since we are
in the Breit frame. We can then calculate the deuteron structure
functions:
\begin{eqnarray}
A&=&G_C^2 + \frac{2}{3} \eta G_M^2 + \frac{8}{9} \eta^2  M_d^4
G_Q^2,\\
\label{eq:A}
B&=&\frac{4}{3} \eta (1 + \eta) G_M^2.
\label{eq:B}
\end{eqnarray}
In terms of $A$ and $B$, the one-photon-exchange interaction yields a
lab-frame differential cross section for unpolarized electron-deuteron scattering~\cite{vOG01}
\begin{equation}
\frac{d \sigma}{d \Omega}=\frac{\sigma_{\rm Mott}}{1 + \frac{2E}{M_d} \sin^2\left(\frac{\theta_e}{2}\right)}
\left[A(Q^2) + B(Q^2) \tan^2\left(\frac{\theta_e}{2}\right)\right].
\label{eq:dcs}
\end{equation}
Here $\theta_e$ is the electron scattering angle and $E$ the electron
energy, and $\sigma_{\rm Mott}$ is the Mott cross section.
Measurement of the differential cross section only yields information
on $G_M$ and a combination of $G_C$ and $G_Q$. A third observable
(usually $t_{20}$, a tensor polarization observable which is sensitive
to $G_Q/G_C$) must be measured if all three form factors are to be
disentangled, and electron-deuteron scattering realize its full
potential as a tool for measuring the deuteron's four-current.

If we make a chiral expansion for the deuteron current
operator~\cite{Phillips:1999am,Walzl:2001vb,Phillips:2003jz} the
results, up to corrections suppressed by three powers of the $\chi$PT
expansion parameter $P \equiv \frac{p,m_\pi}{\Lambda}$, can be written
as:
\begin{eqnarray}
\langle {\bf p}'|J_0({\bf q})|{\bf p}\rangle&=&
|e|\,G_E^{(s)}(Q^2)\, \delta^{(3)}(\bf{p}' - \bf{p} - \bf{q}/2) 
\label{eq:J0},\\
\langle {\bf p}'|{\bf J}({\bf q})|{\bf p} \rangle&=&\left[|e| \frac{{\bf p} + {\bf
    q}/4}{M} G_E^{(s)}(Q^2) + i \mu_S {\bf \sigma}\times {\bf q} G_M^{(s)}(Q^2)\right]
\delta^{(3)}(\bf{p}' - \bf{p} - \bf{q}/2) 
, 
\label{eq:Jplus}
\end{eqnarray}
with $G_E^{(s)}$ and $G_M^{(s)}$ the isoscalar form factors of the
nucleon, and $\mu_S$ the isoscalar nucleon magnetic moment.  In
obtaining these equations we have dropped corrections to $J_\mu$ that
have coefficients $1/M^2$. We will briefly discuss such
``relativistic'' corrections to the deuteron current operator in
Sec.~\ref{sec-higho}.

Sandwiching the operators (\ref{eq:J0}) and (\ref{eq:Jplus}) between
deuteron states yields the following co-ordinate-space integrals for
the deuteron form factors \cite{GG01}:
\begin{eqnarray}
G_C(Q^2)&=&G_E^{(s)}(Q^2) \int d\hbox{r} \left[u^2(r) + w^2(r)\right]
j_0\left(\frac{|{\bf q}| r}{2} \right),\label{eq:GCr}\\
G_Q(Q^2)&=&G_E^{(s)}(Q^2) \frac{6 \sqrt{2}}{Q^2} \int d\hbox{r}
\left[u(r) w(r) - \frac{w^2(r)}{\sqrt{8}}\right] j_2\left(\frac{|{\bf
q}| r}{2} \right),\label{eq:GQr}\\ \frac{2 M}{M_d} G_M(Q^2)&=&G_E^{(s)}(Q^2)
\frac{3}{2} \int d\hbox{r} w^2(r) \left[j_0\left(\frac{|{\bf q}| r}{2}
\right) + j_2 \left(\frac{|{\bf q}| r}{2}\right)\right]\nonumber\\ &&
\qquad + G_M^{(s)}(Q^2) 2 \int d\hbox{r} u^2(r) j_0 \left(\frac{|{\bf
q}| r}{2} \right)\nonumber\\ && + G_M^{(s)}(Q^2) \left\{\sqrt{2}\int d\hbox{r}
u(r) w(r) j_2\left(\frac{|{\bf q}| r}{2} \right) - \int d\hbox{r}
w^2(r) \left[j_0\left(\frac{|{\bf q}|r}{2} \right) - j_2
\left(\frac{|{\bf q}| r}{2}\right)\right] \right\}, \nonumber\\
\label{eq:GMr}
\end{eqnarray}
where $j_0$ and $j_2$ are spherical Bessel functions.

We have demonstrated in Sec.~\ref{sec-randpspace} that we can
determine regulator-independent deuteron wave functions in co-ordinate
space.  They may be used to compute the integrals
Eqs.~(\ref{eq:GCr})--(\ref{eq:GMr}).  At the same time, for any finite
value of $\Lambda$, we can evaluate the form factors $G_C$, $G_Q$, and
$G_M$ using the wave functions obtained by solution of
Eqs.~(\ref{eq:LOV}) and (\ref{eq:schroedinger}). 
In the following, we present the results as a function of
$|{\bf q}|=\sqrt{Q^2}$ following the convention of
Ref.~\cite{Phillips:2003jz}. We restrict ourselves to $|{\bf q}|$
below 1~GeV. This should cover essentially the whole range where
$\chi$PT is expected to converge.

We do this within the context of a chiral expansion for ratios of
deuteron and nucleon form factors, e.g. $G_C/G_E^{(s)}$, so that we
can focus on the predictions of $\chi$ET for deuteron structure. Data
for the single-nucleon form factors could then be employed to
generate final results for deuteron form factors. But, up to the
chiral order to which we work, this procedure is equivalent to using
phenomenological input for $G_E^{(s)}$ and $G_M^{(s)}$ in the
evaluation of expressions for $G_C$, $G_M$, and $G_Q$, e.g. those in
Eqs.~(\ref{eq:GCr})--(\ref{eq:GMr}). Therefore we simply use the
phenomenological single-nucleon form factors of
Ref.~\cite{Belushkin:2006qa} in our evaluation of the deuteron form
factors. These single-nucleon form factors are based on
dispersion relations, and have the practical advantage of quantifying
error bands due to uncertainties in the input experimental data. If we
instead use a strict $\chi$PT expansion for the nucleon form factors
our predictions for deuteron form factors begin to deviate from data
for $|{\bf q}| \sim 300$
MeV~\cite{Walzl:2001vb,Phillips:2003jz}. However, this deviation has
nothing to do with deuteron structure. Instead it is associated with
the restricted range of $|{\bf q}|$ over which $\chi$PT provides a
valid description of nucleon structure. This is the reason that we
employ the single-nucleon form factors of Ref.~\cite{Belushkin:2006qa}
as an input in our calculation. Such a procedure is consistent as regards
our $\chi$ET examination of electron-deuteron scattering because
deuteron and nucleon electromagnetic structure factorize in this
reaction up to corrections that are $O(P^4)$~\cite{Phillips:2003jz}.

We first look at $G_C$ based on the OPE wave functions: see
Fig.~\ref{fig-GCLO}. For $\Lambda \to \infty$, the co-ordinate- and
momentum-space wave functions yield essentially the same result.  The
approach to the $\Lambda \rightarrow \infty$ limit is also shown in
Fig.~\ref{fig-GCLO} and is quite interesting. At $\Lambda=10~{\rm
  fm}^{-1}$ the form factor has converged to the $\Lambda
\rightarrow \infty$ result. However, the result for $\Lambda=3~{\rm
  fm}^{-1}$ predicts a significantly different position of the zero of
$G_C$.  As we shall see below, this feature is particularly affected
by higher-order corrections. The counterterm that would remove this
$\Lambda$ dependence is one such effect, although it is of very high
order: $O(eP^5)$. Small contributions of higher order will have most
pronounced effects at such a zero, where all lower-order effects
cancel by definition, and so sensitivity to the regulator scale in the
vicinity of the minimum is not surprising.  We note that for $|{\bf
  q}| < 600$ MeV, the difference between $\Lambda=3~{\rm fm}^{-1}$ and
$\Lambda=20~{\rm fm}^{-1}$ results is never more than a few per cent.

\begin{figure}[tbp]
\centerline{\includegraphics*[width=80mm,angle=0]{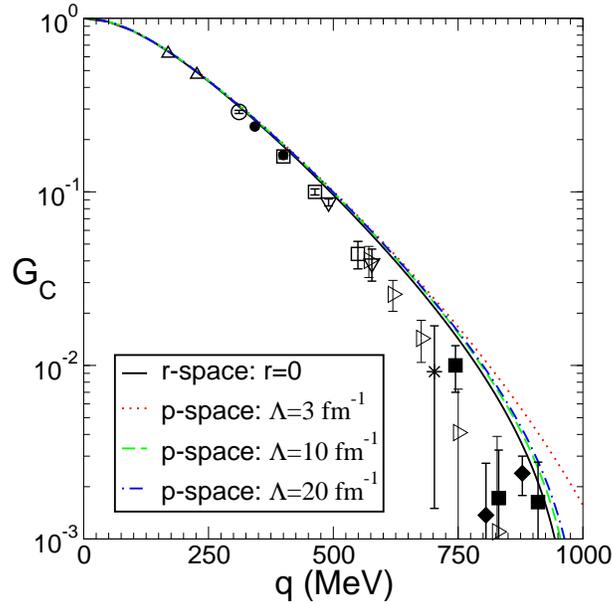}}
\caption{\label{fig-GCLO} Charge form factor of deuterium, $G_C$,
  obtained using different regularizations of the leading-order
  $\chi$ET potential. The red dotted line, green dashed, and blue
  dot-dashed lines are for momentum-space calculations at $\Lambda=3,
  10$, and 20 fm$^{-1}$, respectively. The black solid line is the
  co-ordinate-space result. The experimental data is taken from the
  compilation of Ref.~\cite{Ab00B}: upward triangles represent data
  from the $T_{20}$ measurement of Ref.~\cite{Dm85}, open circle
  \cite{Fe96}, solid circle \cite{Sc84}, open squares \cite{Bo99},
  downward triangles \cite{Gi90}, rightward triangles~\cite{Ni03},
  star \cite{Bo91}, solid squares \cite{Ga94}, solid diamonds
  \cite{Ab00A}.}
\end{figure}

\begin{figure}[tbp]
\centerline{\includegraphics*[width=80mm,angle=0]{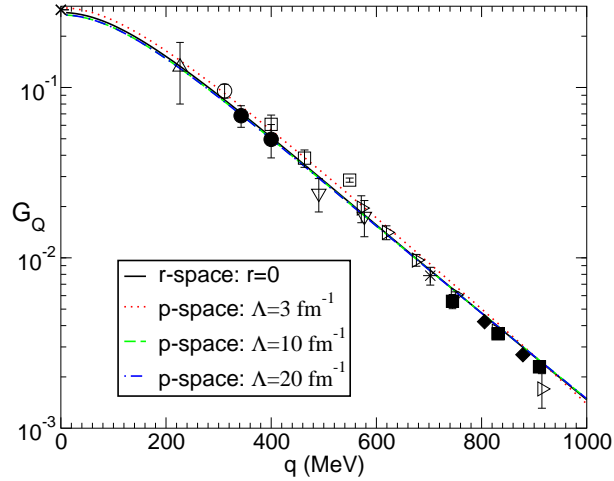}}
\caption{\label{fig-GQLO} Quadrupole form factor of deuterium, $G_Q$,
  obtained using different regularizations of the leading-order
  $\chi$ET potential. Legend for theory curves as in
  Fig.~\ref{fig-GCLO}. Experimental data taken from Ref.~\cite{Ab00B}.}
\end{figure}

\begin{table*}
\caption{\label{tab:deutprop} Cutoff and scheme dependence of the
  kinetic energy $T$ in MeV, the deuteron D-state probability $P_D$ in
  \%, the rms radius $r_d$ in fm, the quadrupole moment $Q_d$ in
  fm$^2$, the asymptotic S-state normalization in fm$^{-1/2}$ and the
  asymptotic S/D ratio $\eta$.  The binding energy was fixed to
  $B=2.225$~MeV in all cases. For wave functions obtained in momentum
  space the cutoff $\Lambda$ is given in fm$^{-1}$ while the r-space
  cut-off $r_c$ is in fm. The first two sets of results are based on
  OPE, while the lower rows are results for TPE (with Set IV $c_i's$).  
Experimental values for observable quantities are listed in the last
row, together with references.}
\begin{center}
\begin{tabular}{c|cc|cccc}
$\Lambda$ &  $T$    &  $P_D$   & $r_d$   & $Q_d$ & $A_S$   & $\eta$   \cr
\hline \hline
2     & 15.43   & 6.98       &  1.90    & 0.311   & 0.845     & 0.0302  \cr
3     & 19.49   & 8.82       &  1.94    & 0.299   & 0.869     & 0.0279   \cr
5     & 28.41   & 7.03       &  1.95    & 0.279   & 0.873     & 0.0263   \cr
10   & 39.62   & 7.29       &  1.94    & 0.277   & 0.869     & 0.0263   \cr
14   & 46.95   & 7.27       &  1.94    & 0.277   & 0.869     & 0.0263   \cr
20   & 57.40   & 7.27       &  1.93    & 0.276   & 0.868     & 0.0263   \cr
\hline
$r$-space &   &            &              &             &               &
              \cr
\hline
1.30  & 10.1 &  10.08      &  2.06    & 0.359   & 0.925     & 0.0302  \cr
1.03  & 16.97 &  8.95      &  1.99    & 0.312   & 0.894     & 0.0279  \cr
0.80  & 25.40 &  8.07      &  1.96    & 0.288   & 0.877     & 0.0268  \cr
0.40  & 33.48 &  7.21      &  1.94    & 0.277   & 0.870     & 0.0263  \cr
0.20  & 61.74 &  7.29      &  1.94    & 0.276   & 0.868     & 0.0263  \cr
0.10  & 89.0  &  7.31      &  1.94    & 0.276   & 0.868     & 0.0263  \cr
\hline \hline
$\Lambda$ &    &     &    &  &   &   \cr  \hline
3.90   &  30.27  & 8.86    & 2.00       & 0.286   &  0.898    & 0.0257   \cr
8.45   &  41.60  & 7.67    & 1.97       & 0.276   &  0.888    & 0.0254   \cr
8.57   &  46.61  & 8.96    & 1.97       & 0.280   &  0.883    & 0.0260   \cr
14.1   &  55.24  & 8.31    & 1.97       & 0.277   &  0.884    & 0.0256   \cr
20.9   &  69.87  & 8.46    & 1.96       & 0.277   &  0.883    & 0.0256   \cr
\hline
$r$-space &   &            &              &             &             
&                \cr
\hline
0.92   &  22.93  & 6.72    & 2.00       & 0.286   &  0.898    & 0.0257   \cr
0.67   &  39.45  & 7.36    & 1.98       & 0.276   &  0.888    & 0.0254   \cr
0.66   &  46.74  & 8.85    & 1.97       & 0.281   &  0.883    & 0.0260   \cr
0.50   &  59.53  & 8.06    & 1.97       & 0.277   &  0.885    & 0.0256   \cr
0.35   &  112.8  & 8.13    & 1.97       & 0.277   &  0.884    & 0.0256   \cr
0.20   &  336.7  & 8.14    & 1.97       & 0.276   &  0.884    & 0.0256   \cr
\hline \hline
expt          & --  &   --       & 1.953(3) \cite{Klarsfeld:1986}  &
0.2859(3) \cite{CR71,BC79,Ericson:1982ei} & 0.8781(44) \cite{Borbely:1985} &
0.0256(4) \cite{Rodning:1990zz}
\end{tabular}
\end{center}
\end{table*}

The results obtained in co-ordinate-space and using large cutoffs in
momentum space also agree very well for $G_Q$. As seen in
Fig.~\ref{fig-GQLO}, convergence 
of the momentum-space results to the $\Lambda
\rightarrow \infty$ limit is even more rapid there. Presumably, this is
because, as was observed in Ref.~\cite{SS01}, $G_Q$ is less sensitive
to short-distance details than is $G_C$. There is, however, one
exception to this: the value of $G_Q$ at $Q^2=0$, i.e. the deuteron
quadrupole moment, has significant evolution with $\Lambda$, dropping
by more than 10\% between $\Lambda=3~{\rm fm}^{-1}$ and
$\Lambda=\infty$. This can be seen in Table~\ref{tab:deutprop}, where,
for completeness, we also compile some other basic properties of our
deuteron wave functions. As was observed in
Ref.~\cite{Phillips:2006im}, such $\Lambda$-dependence is
symptomatic that the $O(eP^5)$ counterterm impacts $Q_d$
more significantly than one would naively expect. A
counterterm of natural size at $O(eP^5)$ can absorb this large
a cutoff dependence, and once it does, the remaining short-distance
effects in $G_Q$ are minimal below $|{\bf q}|=600$~MeV~\cite{Phillips:2006im}.

In $G_M$ the $\Lambda \rightarrow \infty$ limit again yields the
co-ordinate-space result. But the results for low cutoffs deviate from
the $\Lambda \rightarrow \infty$ limit more than in any of the other
form factors, as displayed in Fig.~\ref{fig-GMLO}. They differ by
10\% at $|{\bf q}|=600$~MeV, which corresponds to probing momenta of
$|{\bf q}|/2=300$~MeV in the nucleus itself.  This sensitivity is
entirely consistent with the presence of a counterterm for $G_M$ at
$O(eP^4)$, which is only N$^2$LO. Indeed, our calculation of $G_M$
includes all NLO effects in the deuteron three-current ${\bf J}$
(apart from small $1/M^2$ corrections) so the counterterm would appear
in ${\bf J}$ at the next order beyond what is presented here.

\begin{figure}[tbp]
\centerline{\includegraphics*[width=80mm,angle=0]{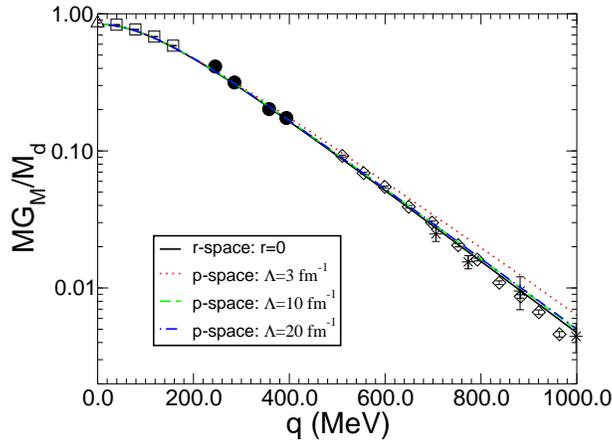}}
\caption{\label{fig-GMLO} Magnetic form factor of deuterium $G_M$ obtained
  using different regularizations of the leading-order $\chi$ET
  potential. Legend for theory curves as in Fig.~\ref{fig-GCLO}. 
Experimental data from
deuteron magnetic moment, open triangle~\cite{Li65}; the
parameterization of Ref.~\cite{Si01}, open squares; and measurements
of $B(Q^2)$: solid circles~\cite{Si81}, open diamonds~\cite{Au85}, and
stars~\cite{Cr85}.}
\end{figure}

These issues aside, we see that the LO result for all three form
factors is in remarkably good agreement with the data.  In fact, as
$\Lambda \rightarrow \infty$, the agreement with the data improves
consistently. Obviously, the unphysical, spurious bound states affect
the structure of the wave functions only at distances which are not
relevant for momentum transfers $|{\bf q}|$ below 1~GeV.  
The spurious bound states in the
$^3$S$_1$-$^3$D$_1$ channel have no impact on these observables in the
domain of validity of $\chi$ET.  Consequently their appearance in the
spectrum in no way signals a breakdown of this approach.

The agreement obtained between the data and the form factors evaluated
with LO wave functions is surprising and also encouraging. 
One advantage of $\chi$ET over phenomenological models of deuterium is
its ability to systematically improve its predictions. Therefore in
the next section we examine what happens when certain sub-leading
effects are included in the wave functions and also discuss the impact
on the predictions of pieces of the current operator that are of higher
chiral order.

\section{Impact of two-pion exchange and exchange currents on the form factors}
\label{sec-higho}

To begin our discussion of higher-order effects on the deuterium form
factors we first discuss some of the higher-order contributions to the
current operators.  Both $J_0$ and ${\bf J}$ have a chiral expansion,
and to produce the results displayed in the previous section we have
ignored terms of order $\left(\frac{P}{\Lambda}\right)^3$. We have
also dropped terms suppressed by $\left(\frac{P}{M}\right)^2$. In
general corrections $\left(\frac{P}{\Lambda}\right)^n$ will be larger
than corrections $\left(\frac{P}{M}\right)^n$, as emphasized in
Weinberg's original power counting used, e.g., in
\cite{Ordonez:1995rz,Ep05}.  $\left(\frac{P}{\Lambda}\right)^n$
modifications to Eqs.~(\ref{eq:GCr})--(\ref{eq:GMr}) are associated
with two-body operators because we have used phenomenological nucleon
form factors. The strength of these terms is not fixed by any
symmetry and must be determined by data (from either the
single-nucleon or two-nucleon sector).  While the first such effects
in $G_M$ are of relative order $P^3$, the first occurrence of such a
two-body operator in $G_C$ and $G_Q$ is not until relative order
$P^4$. It is therefore higher-order than all the TPE effects in $V$
that we will consider here.

On the other hand, corrections to Eqs.~(\ref{eq:GCr}) and
(\ref{eq:GQr}) of nominal size $\left(\frac{P}{M}\right)^n$ occur in
$J_0$ with $n=2$. These are corrections to the one-body part of $J_0$,
and they can be obtained by demanding that matrix elements of the
one-body current operator obey the correct transformation properties
for Lorentz boosts to frames with velocities $v \ll c$.  If we do this,
we find a modification to the one-body part of $J_0$ that introduces
an additional factor of $1-\frac{Q^2}{8M^2}$ in Eq.~(\ref{eq:J0}). In
addition, if we are to have low-energy Lorentz covariance for {\it
  deuteron} matrix elements of the current operator $(J_0,{\bf J})$ the
momentum transfer $|{\bf q}|$ that is used in evaluating the integrals
of Eq.~(\ref{eq:GCr})--(\ref{eq:GMr}) must also be redefined. (This
accounts for the boost of the deuteron wave function from the deuteron
center-of-mass to the Breit frame~\cite{AA96,Phillips:2003jz}.) 
Both of these are small effects at the range of $Q^2$ values considered here. 
Specifically, the former
does not affect the minimum of $|G_C|$ at all, while the latter shifts
it to the right by 3\%, i.e. about 15 MeV for the LO results presented
above. A two-nucleon effect in $J_0$ that is of order
$\frac{P^3}{\Lambda^2 M}$ will be discussed explicitly below.

Because these $1/M^2$ effects to the operator $J_0$ are generically
small, in what follows we focus on the impact of improving the wave
functions. In particular, we first want to apply the NNLO wave
functions that were discussed in Sec.~\ref{sec-randpspace} and
which incorporate corrections of size
$\left(\frac{P}{\Lambda}\right)^3$.  The procedure used to evaluate
$G_C$, $G_Q$, and $G_M$, with these wave functions is the same as in
Sec.~\ref{sec-LOFF}, i.e. single-nucleon form factors of Belushkin,
Hammer, and Mei{\ss}ner are incorporated in the calculation to account
for the effects of single-nucleon structure as per
Eqs.~(\ref{eq:GCr})--(\ref{eq:GMr}).

The pattern of convergence to the $\Lambda \rightarrow \infty$ limit
is very much the same as in the case of the OPE potential discussed in
Sec.~\ref{sec-LOFF}.  Here, we first show results for $G_C$ and $G_Q$,
in Figs.~\ref{fig-GQTPEcvgce} and \ref{fig-GCTPEcvgce},
respectively. $G_Q$ remains largely insensitive to $\Lambda$.  At
$|{\bf q}|=0$, the sensitivity is as large as that at any $|{\bf q}|$, and is
approximately 3\% over the range from $\Lambda=3.9~{\rm fm}^{_1}$ to
$\Lambda=14.1~{\rm fm}^{-1}$. Perhaps not coincidentally, this is
roughly the size of the correction needed to move the value of $Q_d$
obtained with NNLO $\chi$ET to the experimental
result~\cite{Phillips:2006im}.  We find that $G_C$ is more sensitive
to $\Lambda$.  In particular, the cutoff dependence of $G_C$ around
its zero is somewhat larger than naive expectations given the
breakdown scale of $\chi$PT, and so we stress that that such
expectations might be misleading, especially in the vicinity of a zero
like this.  For both form factors, the limit as $\Lambda \rightarrow
\infty$ of the results with the momentum-space cutoff is equal (within
numerical uncertainties) to the result obtained via imposing a
boundary condition in co-ordinate space at $r=0$.

In Fig.~\ref{fig-GQTPEcvgce} we also show results for one of the
modern $\chi$PT $NN$ interactions of Ref.~\cite{Ep05}.  For this
observable, we will show below that the uncertainties due to different
sets of $\pi$N LECs, the nucleon form factor and different choices for
$\eta$ are rather small. But Fig.~\ref{fig-GQTPEcvgce} shows that
results based on the NNLO $\chi$ET wave functions of Ref.~\cite{Ep05},
which employ a regulator $\exp(-p^6/\Lambda^6)$ with $\Lambda$ in the
range 500--700 MeV, do not overlap with the predictions from our wave
functions at any $|{\bf q}|$.

\begin{figure}[tbp]
\centerline{\includegraphics*[width=80mm,angle=0]{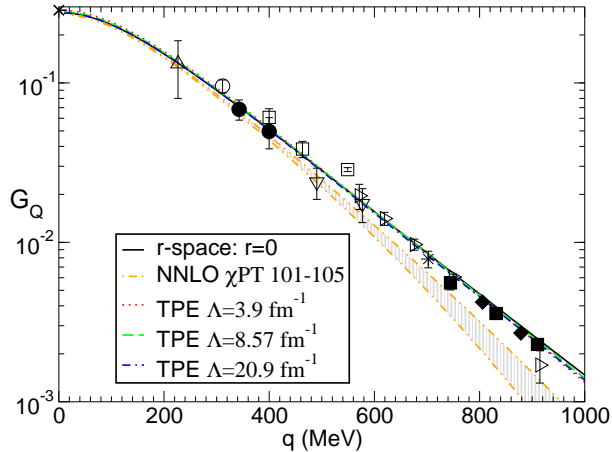}}
\caption{\label{fig-GQTPEcvgce} Quadrupole form factor of deuterium $G_Q$ 
  obtained using
  using different regularizations of the TPE
  potential. Legend as in Fig.~\ref{fig-GCLO}, with slight differences
  in cutoffs indicated.}
\end{figure}

\begin{figure}[tbp]
\centerline{\includegraphics*[width=80mm,angle=0]{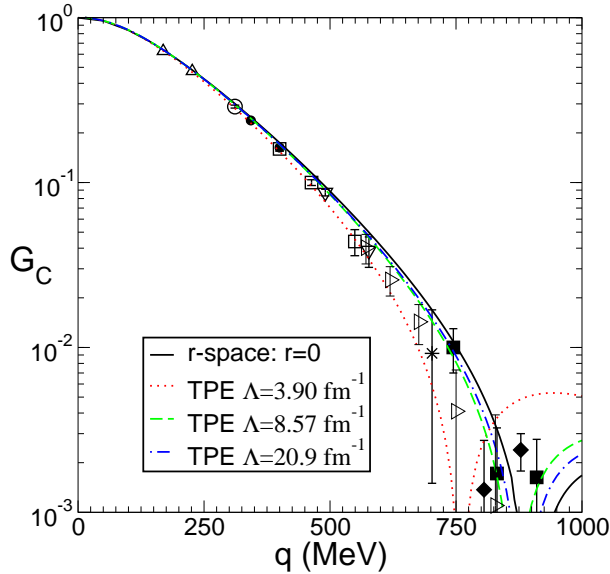}}
\caption{\label{fig-GCTPEcvgce} Charge form factor of deuterium $G_C$ 
  obtained using
  using different regularizations of the TPE
  potential. Legend as in Fig.~\ref{fig-GCLO}, with slight differences
  in cutoffs indicated.}
\end{figure}

Having demonstrated the equivalence of co-ordinate- and momentum-space
formulations of TPE in the limit in which the regulator is removed, we
now examine the impact of TPE on the form factors themselves by
comparing results for the various sets of $c_i$'s given in
Table~\ref{tab:cival}.  We begin with results for $G_C$, which are
presented in Fig.~\ref{fig-GC-exp}, where we also compare to data.
Here we display the results for Sets I, III, and IV. (The results
for Set II are very similar to those obtained with Set I.)  The error
band shown for the Set IV choice incorporates both uncertainties in
the deuteron wave function (mainly those from varying the asymptotic
$D/S$-ratio in this case), as well as the uncertainty bands in the
single-nucleon form factors quoted in
Ref.~\cite{Belushkin:2006qa}. Sensitivity of $G_C$ to these
uncertainties is similar for the other two sets of $\pi$N LECs. We
note that such uncertainties are smaller than the sensitivity to
different choices for the $c_i$'s.

The results clearly show that TPE corrections are small, as
anticipated from the chiral expansion, and that---independent of the
$\pi$N LECs used to compute TPE effects---they shift the minimum of
$|G_C|$ to the left.  This yields a rather remarkable agreement
between the $G_C$ obtained once TPE is included in the deuteron wave
function and experimental data.  We will see below that effects of
similar size can be expected from exchange-current contributions to
the operator $J_0$. Therefore, at this point, no further conclusions
on the size of LECs in TPE can be drawn. However, the results clearly
indicate that it will be interesting to study $G_C$ again taking such
two-body effects in $J_0$ into account.  The results of
Fig.~\ref{fig-GC-exp} also suggest that the minimum of $|G_C|$ could
be a fruitful place to look for improvements in the EFT's description
of data due to the inclusion of explicit $\Delta(1232)$ degrees of
freedom. Such calculations are, however, beyond the scope of this
study.

\begin{figure}[tbp]
\centerline{\includegraphics*[width=80mm,angle=0]{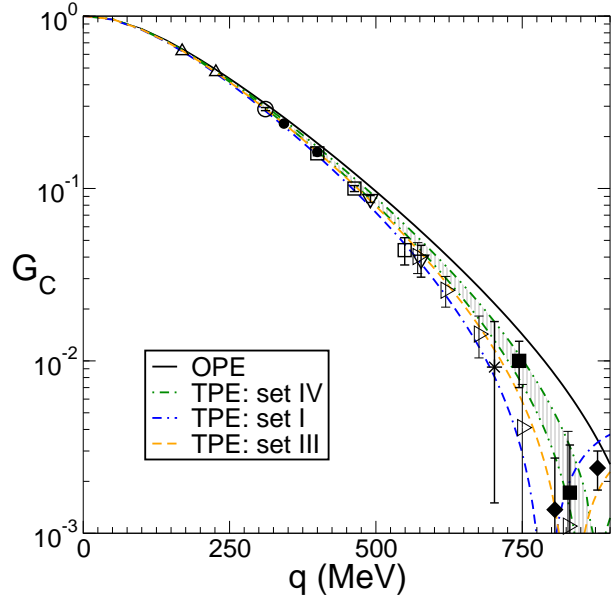}}
\caption{\label{fig-GC-exp} Comparison of charge form factor $G_C$
  with (blue, green, and orange broken curves) and without (black
  solid curve) the NNLO corrections included in the $NN$
  potential. The green double-dot-dashed, blue double-dash-dotted, and
  orange short-dashed curves use the different choices of the $c_i$'s
  listed in Table~\ref{tab:cival}.  The error bands in the theoretical
  calculation shown for the ``Set IV'' choice of $\pi$N LECs
  incorporate the experimental uncertainties in both the input value
  of $\eta$ and the nucleon form factors of
  Ref.~\cite{Belushkin:2006qa}. Data from the compilation of
  Ref.~\cite{Ab00B}.}
\end{figure}

\begin{figure}[tbp]
\centerline{\includegraphics*[width=80mm,angle=-90]{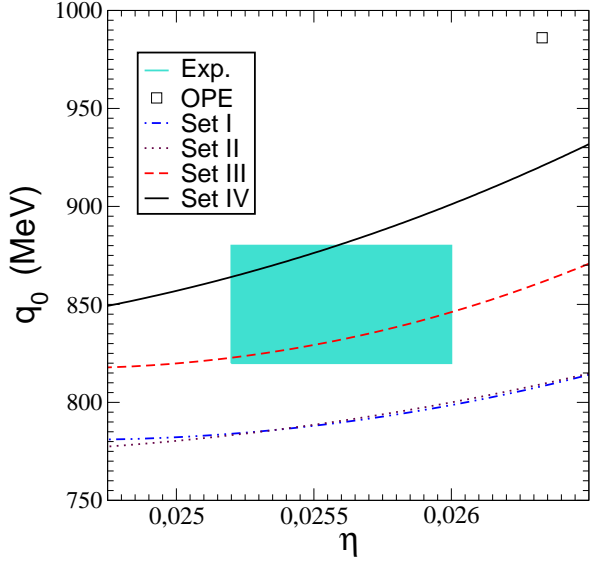}}
\caption{\label{fig-minplot} The position of the zero in the charge
  form factor, $q_0$, as a function of the asymptotic D-to-S ratio
  $\eta$ for the four different sets of $\pi$N LECs considered here.
  The (1$\sigma$) experimental constraints on $\eta$ and $q_0$ are
  indicated  by the shaded region. The result found for $\eta$ and
  $q_0$ from the LO $\chi$ET potential with $\Lambda \rightarrow
  \infty$ is given by the square.}
\end{figure}

If the NNLO potential is used to compute deuteron wave functions the
deuteron's asymptotic D-to-S ratio, $\eta$, is a free parameter of the
calculation. In Fig.~\ref{fig-minplot} we show how the position of
$|G_C|$'s minimum depends on $\eta$. Note that in order to make such a
plot we need to specify the long-range potential, i.e. the set of
$c_i$'s we used for TPE. Also shown in Fig.~\ref{fig-minplot} is the
regulator-independent result obtained with the OPE potential, where
$\eta$ is {\it not} a free parameter, but is determined by the
dynamics.  This figure demonstrates that the position of the $|G_C|$
minimum is only weakly dependent on $\eta$. In particular, had we used
$\eta_{\rm OPE} (\gamma ) \approx 0.026333$ as our input instead of
the central experimental value $\eta_{\rm expt}=0.0256$ the minimum
would only have shifted about 4\% to the right. But the overall shift
in the minimum from the OPE result to the TPE result with Set IV
$c_i$'s is 12\% (the shifts for Set I-III $c_i$'s are even
larger). So roughly two-thirds of the shift in the $|G_C|$
minimum arises from the inclusion of TPE and corresponding changes of
the wave function at distances $r \sim 1/m_\pi$. When
regulator-independent wave functions are employed two-pion exchange
has a significant impact on the $|{\bf q}|$ at which the zero of $G_C$
occurs, regardless of which $c_i$'s are used in its evaluation.

Turning our attention now to Fig.~\ref{fig-GQ-exp}, we see that the
OPE and TPE results for $G_Q$ are very
close together. Above $|{\bf q}|=350$ MeV the effect of TPE is
to shift $G_Q$ downwards, irrespective of the set of $c_i$'s
chosen. However, the error bands on the OPE and TPE 
calculations are always overlapping (recall that though we
only show the error band for the TPE: Set IV calculation, all three
TPE calculations have bands of similar size due to uncertainties in
their input). The only place where an unambiguous difference between
the different $NN$ potentials can be seen is in the quadrupole
moment. But here it is difficult to draw any definitive conclusion,
since all calculations have central values that under-predict the
experimental result $Q_d=0.2859(3)~{\rm fm}^2$~\cite{CR71,BC79}. This
discrepancy is, however, consistent in all cases with the expected 
short-distance contribution to $G_Q$ which occurs
at $O(e P^5)$ in the chiral expansion for $J_0$.  All of this suggests
that $G_Q$ is not a good place to attempt to test the TPE 
contribution to the deuteron wave function.

\begin{figure}[tbp]
\centerline{\includegraphics*[width=80mm,angle=0]{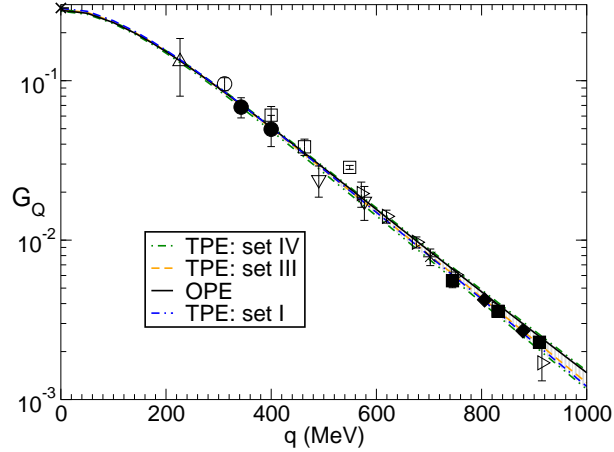}}
\caption{\label{fig-GQ-exp} Comparison of quadrupole form factor $G_Q$
  with (blue, orange, and green broken curves) and without (black
  solid curve) the NNLO corrections included in the $NN$
  potential. The green double-dot-dashed, blue double-dash-dotted, and
  orange short-dashed curves use the different choices of the $c_i$'s
  listed in Table~\ref{tab:cival}.  The error bands in the theoretical
  calculation shown for the ``Set IV'' choice of $\pi$N LECs
  incorporate the experimental uncertainties in both the input value
  of $\eta$ and the nucleon form factors of
  Ref.~\cite{Belushkin:2006qa}. Data from the compilation of
  Ref.~\cite{Ab00B}.}
\end{figure}

In contrast, $G_M$ is quite sensitive to the choice of dynamics for
the long-range part of the $NN$ potential, as is seen in
Fig.~\ref{fig-GM-exp}. The uncertainties from the input value of
$\eta$ and the single-nucleon form factors are sizeable, but even
allowing for these uncertainties there are differences between OPE and
TPE wave functions at the 1--2$\sigma$ level. And at least at $|{\bf q}| >
0.5~{\rm GeV}$, there is also significant sensitivity to the choice
of $\pi$N LECs that is employed in the sub-leading TPE. Unfortunately
the effect of TPE here is to worsen the excellent agreement with data
that is achieved with OPE. (Although we note that Set IV is still
marginally consistent with data if we allow for all input
uncertainties.) However, it should be remembered that the
theoretical uncertainty due to higher-order corrections in $G_M$ is
not depicted here. Such effects in the current ${\bf J}$ are suppressed by
$P^3/\Lambda^3$ relative to leading, the same relative size as
TPE, and so should be included if a consistent
calculation up to that order of $G_M$ in the chiral expansion is
desired. Until such a calculation is completed only tentative
conclusions about the impact of the $O(P^3)$ pieces of the $NN$
potential on $G_M$ can be drawn. 

\begin{figure}[tbp]
\centerline{\includegraphics*[width=80mm,angle=0]{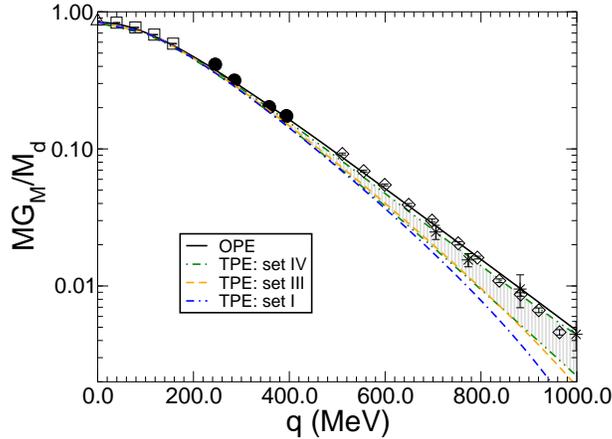}}
\caption{\label{fig-GM-exp} Comparison of magnetic form factor $G_M$
  with (blue, orange, and green broken curves) and without (black
  solid curves) the NNLO corrections included in the $NN$ potential.
  The green double-dot-dashed, blue double-dash-dotted, and orange
  short-dashed curves use the different choices of the $c_i$'s listed
  in Table~\ref{tab:cival}.  The error bands in the theoretical
  calculation shown for the ``Set IV'' choice of $\pi$N LECs
  incorporate the experimental uncertainties in both the input value
  of $\eta$ and the nucleon form factors of
  Ref.~\cite{Belushkin:2006qa}. Data as in Fig.~\ref{fig-GMLO}.}
\end{figure}

Similar caution is advisable in the interpretation of our results for
$G_C$ and $G_Q$. In this case there is a two-body piece of $J_0$ at
relative order $P^3$ whose coefficient is fixed by the low-energy
consequences of Lorentz covariance. This operator was first derived in
Ref.~\cite{Ri84}, and was re-derived in $\chi$ET in
Ref.~\cite{Phillips:2003jz}. The two-body matrix element of this piece
of $J_0$ can be computed reliably with the momentum-space wave
functions of Section~\ref{sec-randpspace}. It converges to a definite
result as $\Lambda \rightarrow \infty$ and shifts the minimum of
$|G_C|$ to the left by about 100 MeV, worsening the agreement with
data seen in Fig.~\ref{fig-GC-exp}. However, as emphasized by
Friar~\cite{Fr80} and Adam and Arenh\"ovel~\cite{Ad93}, this
correction to $J_0$ is associated with $P^2/M^2$ corrections to the
OPE potential used in generating the wave functions.  None of those
$P^2/M^2$ pieces of $V$ were included in our calculations. Therefore a
full evaluation of the relevant $1/M$-suppressed effects remains to be
performed with these wave functions. Such an evaluation is, however,
beyond the scope of this paper.

\section{Conclusion}
\label{sec-concl}

Electron-deuteron scattering provides a window onto $NN$ dynamics that
gives complementary information to that obtained in $NN$
scattering. It allows us to work at fixed total energy of the $NN$
system and examine the electromagnetic response of the system as a
function of the momentum-transfer squared, $q^2$ (which is equal to
$-|{\bf q}|^2$ in the Breit frame). In 
$\chi$ET the predictions for the three electromagnetic form factors
$G_C$, $G_Q$, and $G_M$ that determine all elastic electron-deuteron
scattering observables in the one-photon-exchange approximation are
given solely by single-nucleon operators up to corrections of relative
order $P^3$. Consequently the venerable formulae
(\ref{eq:GCr})--(\ref{eq:GMr}) for these form factors as Bessel
transforms of probability densities are valid to this level of accuracy.

Examination of these formulae indicates that the charge form factor is
sensitive to distances up to a value of $r$ given by
\begin{equation}
r \approx \pi{\Big/}\frac{|{\bf q}|}{2},
\end{equation}
where the factor of $\pi$ upstairs arises from the first zero of
$j_0(x)$ and the factor of $1/2$ in the momentum transfer occurs
because only half of ${\bf q}$ is transmitted to the relative degree of
freedom. Thus as we move through the momentum range $0 \leq |{\bf q}|
\leq 1$ GeV $G_C$ changes from an average of the probability density
over all $r$ to one where the important dynamics is that taking place
at $r \approx 1.2$ fm. This radius is well inside that at which OPE
is active, and indeed is small enough that some TPE effects are probed. 
Once $|{\bf q}| \approx 1$ GeV we expect that
two-body operators which scale as $\left(\frac{|{\bf
    q}|}{\Lambda_{\chi {\rm SB}}}\right)^n$ with $n \geq 3$ will no
  longer be suppressed, and so the dominance of the single-nucleon
  contributions to the deuteron four-current is no longer guaranteed
  there. But the momentum range $|{\bf q}| \leq 1$ GeV already
  provides a wide kinematic domain over which we can test $\chi$ET
  predictions for the deuteron wave functions (or equivalently---up to
  $P^3$ corrections---form factors), and examine the impact of
  regularization on these predictions.

Several calculations of these form factors that use $\chi$ET already
exist in the 
literature~\cite{Phillips:1999am,Walzl:2001vb,Phillips:2003jz,Phillips:2006im}. But here we focused particularly
on the possibility to obtain regulator-independent predictions for
$G_C$, $G_Q$, and $G_M$ in the regime $|{\bf q}| \leq 1$ GeV.  We
found that the results for all three form factors become stable as the
momentum-space cutoff on the $\chi$ET calculation, $\Lambda$, is taken
to infinity. The resulting predictions are therefore free of artifacts
due to the particular function that is chosen as a short-distance
regulator. They can also be obtained by employing a co-ordinate-space
wave function where the Schr\"odinger equation is solved for all $r >
0$ using $NN$ potentials derived from chiral perturbation theory. We
found that the momentum-space cutoff must be taken to be more than
2 GeV before stability is obtained in some cases (e.g. $G_C$ near its
zero, $G_M$ at $|{\bf q}| \approx 0.45~{\rm GeV}$).  We therefore caution
that invoking naive-dimensional-analysis estimates to set the size of
cutoffs that should be employed in $\chi$ET can lead to the presence
of significant cutoff artifacts in the results for observables.

We were able to perform these calculations for both the LO $NN$
potential (which consists solely of one-pion exchange at long range)
and for the NNLO $NN$ potential. Because both of these potentials are
singular and attractive as $r \rightarrow 0$ they can generate a
shallow (deuteron) bound state. But their singular nature means that
as $\Lambda \rightarrow \infty$ they also generate spurious bound states at
energies beyond the range of applicability of the theory. However,
these spurious bound states have no impact on the deuteron form
factors, and we found no indication that their appearance signals a
breakdown of the $\chi$ET approach.

We were not able to perform a calculation with the NLO $\chi$PT $NN$
potential, since the potential at this order is singular and
repulsive. In consequence it is impossible to take the $\Lambda
\rightarrow \infty$ limit with this $V$ inserted in the Schr\"odinger
equation and still have a spectrum with a shallow bound state. An
order-by-order assessment of the convergence of $\chi$ET must
therefore await the development of a power counting which does not
have this deficiency. In this context we note that the version of
$\chi$PT in which the $\Delta(1232)$ is included as an explicit degree
of freedom leads to a potential at NLO in the (modified) chiral
expansion that is singular and attractive. Indeed, in the
${}^3$S$_1$--${}^3$D$_1$ channel the NLO potential in the theory with
explicit Deltas is very similar to the NNLO potential we have used
here. Therefore we believe that a study employing the NLO $V$ from
$\chi$PT with explicit Deltas---while definitely called for---will
likely reach similar conclusions to those  found here using the NNLO $V$
from standard (Delta-less) $\chi$PT.
        
The calculations to LO in $\chi$ET gave predictions for deuteron
electromagnetic structure which are remarkably close to experiment for
$|{\bf q}| \leq 0.6~{\rm GeV}$ once the $\Lambda \rightarrow \infty$
limit was taken. The only significant disagreement between the LO
prediction and the data in the compilation of Ref.~\cite{Ab00B}
occurred in the vicinity of the zero of $G_C$. But even there the
discrepancy between theory and data was within the expected
$\left(\frac{Q^2}{4 \Lambda_{\chi \rm SB}^2} \right)$, which is about
25\% at 1 GeV. 

The corrections to the LO result once the NNLO potential was employed
to compute deuteron structure were also consistent with this
expectation, proving to be small, especially in the case of $G_Q$,
which is already well described by the LO calculation. Our results
also show that the discrepancy between the LO $\chi$ET calculation and
experimental data for $G_C$ in the vicinity of its zero may be
eliminated when such a higher-order calculation is performed. In
particular, we showed that the position of the minimum in $G_C$ is
sensitive to details of the two-pion-exchange interaction that is
present in the NNLO $\chi$PT potential. This will be an interesting
point to focus on in future investigations of two-pion exchange.  In
contrast, the good description of $G_M$ in LO appears to be
accidental. Various small but significant higher-order effects
apparently cancel for this observable, but this needs to be checked in
a complete higher-order calculation.

We also assessed the impact of the $O(P^3)$ corrections to $J_0$ on
our results for $G_C$ and found that they converge to a definite limit
as $\Lambda \rightarrow \infty$. Although we did not include any
$\chi$PT corrections to the leading-order $J_\mu$
in the results we presented, our estimates of the $O(P^3)$ effects in
$J_0$ show that the exchange-current corrections have roughly the same
impact on $G_C$ in the vicinity of its zero as do the $O(P^3)$
contributions to $V$. This suggests that the power counting for the
potential and the currents is working well in this domain.

All of this makes it very interesting to complete consistent
higher-order calculations of deuteron form factors using these
regulator-independent wave functions. To do this the $O(P^3)$ pieces
of $J_\mu$ will have to be computed fully and in a manner that is
consistent with the treatment of $1/M^2$ corrections in $V$. Only when
such a computation is performed will we know if the improvement found
here with the NNLO $\chi$ET deuteron wave functions represents
genuinely good convergence of the chiral expansion for these
observables or not. The success we observed here may be merely a
fortunate result of examining only the corrections to $V$.  And an
assessment of the impact that $O(P^4)$ (and higher) pieces of $J_0$
will also be necessary if a definitive conclusion as regards the
impact of TPE contributions on the position of $G_C$'s zero is to be
reached. Irrespective of these issues, though, it is clear that the
deuteron charge form factor is sensitive to details of the chiral
dynamics that is at work in deuterium.

\begin{acknowledgement}
We are grateful to the organizers of the Trento workshop on ``QCD and
Nuclear Forces: Never the Twain Shall Meet?'' (June 2005) for
providing a stimulating environment which enabled us to come together
and initiate this research. We thank Maxim Belushkin for providing the
nucleon form factor data files, and Evgeny Epelbaum, Ulf Mei\ss ner and 
Matthias Schindler for comments on the manuscript.

The work of DP was supported under US Department of Energy grant
DE-FG02-93ER40756.  The work of ERA  is supported in part by funds
provided by the Spanish DGI and FEDER funds with grant
no. FIS2005-00810, Junta de Andaluc{\'\i}a grants no.  FQM225-05, EU
Integrated Infrastructure Initiative Hadron Physics Project contract
no. RII3-CT-2004-506078.  MPV has been funded by the Deutsche
Forschungsgemeinschaft (SFB/TR 16), Helmholtz Association (contract
number VH-NG-222). Part of the numerical calculations have been 
performed on JUMP and JUBL of the JSC in J\"ulich, Germany. 
\end{acknowledgement}

\end{onecolumn}

\end{document}